\documentclass[english,prd,twocolumn,superscriptaddress,nofootinbib,preprintnumbers]{revtex4}
\usepackage[latin1]{inputenc}
\usepackage{graphicx}
\usepackage{amssymb}

\usepackage{amsfonts}
\usepackage{dcolumn}
\usepackage{hyperref}

\def\be{\begin{equation}}
\def\ee{\end{equation}}
\def\ba{\begin{eqnarray}}
\def\ea{\end{eqnarray}}
\def\bs{\begin{subequations}}
\def\es{\end{subequations}}

\usepackage{color}

\newcommand{\GB}{\mathcal{G}}
\newcommand{\de}{\mathrm{d}}

\begin{document}

\title{Matter instabilities in general Gauss-Bonnet gravity}

\author{Antonio De Felice}
\affiliation{Department of Physics, Faculty of Science, Tokyo University of Science, 
1-3, Kagurazaka, Shinjuku-ku, Tokyo 162-8601, Japan}

\author{David F. Mota}
\affiliation{Institute of Theoretical Astrophysics 
University of Oslo 0315, Oslo Norway}

\author{Shinji Tsujikawa}
\affiliation{Department of Physics, Faculty of Science, Tokyo University of Science, 
1-3, Kagurazaka, Shinjuku-ku, Tokyo 162-8601, Japan}

\begin{abstract}
  We study the evolution of cosmological perturbations in $f(\GB)$
  gravity, where the Lagrangian is the sum of a Ricci scalar
  $R$ and an arbitrary function $f$ in terms of a Gauss-Bonnet term
  $\GB$. We derive the equations for perturbations assuming matter 
  to be described by a perfect fluid with a constant equation of state $w$. 
  We show that density perturbations in perfect fluids
  exhibit negative instabilities during both the radiation and 
  the matter domination, irrespective of the form of $f(\GB)$.
  This growth of perturbations gets stronger on smaller
  scales, which is difficult to be compatible with the observed galaxy
  spectrum unless the deviation from General Relativity is very 
  small. Thus $f(\GB)$ cosmological models are effectively ruled out
  from this Ultra-Violet instability, even though they can be
  compatible with the late-time cosmic acceleration and local gravity
  constraints.
\end{abstract}

\date{\today}

\maketitle

\section{Introduction}

Independent observational evidence for dark energy has
stimulated the idea that General Relativity (GR) may be modified on
large distances to give rise to a late-time cosmic acceleration
\cite{review}.  A simple dark energy scenario constructed in this vein
is so-called $f(R)$ gravity in which $f$ is a function of the Ricci
scalar $R$ \cite{fRori}.  Although there are some restrictions to the
functional form of $f(R)$ to satisfy both cosmological and local
gravity constraints, it is possible to design viable models
\cite{fRviable} that can be distinguished from GR at least in the
metric formalism of $f(R)$ gravity.

The $f(R)$ gravity in the metric formalism corresponds to the
so-called Brans-Dicke theory with a parameter $\omega_{\rm BD}=0$ in
the presence of a potential of gravitational origin \cite{Chiba}.  One
can generalize this to scalar-tensor theories with an arbitrary
Brans-Dicke parameter $\omega_{\rm BD}$. In fact it is possible to
construct scalar-field potentials that can be responsible for the
cosmic acceleration, while at the same time satisfying local gravity
constraints \cite{TUMTY}.  These models, including $f(R)$ gravity,
exhibit several interesting observational signatures such as the
phantom equation of state \cite{AT07}, the modified matter power
spectrum \cite{fRmatter}, and the modified weak lensing 
spectrum \cite{WL}.

The Ricci scalar $R$ is not the only scalar quantity which is used to
change gravity, as we can easily construct other scalar quantities
such as $R_{\mu \nu}R^{\mu \nu}$ and $R_{\mu \nu \rho \sigma}R^{\mu
  \nu \rho \sigma}$ from the Ricci tensor $R_{\mu \nu}$ and the
Riemann tensor $R_{\mu \nu \rho \sigma}$ \cite{PQR}.  However, for the
Gauss-Bonnet (GB) curvature invariant
\begin{equation}
\GB \equiv R^2-4R_{\mu \nu}R^{\mu \nu}
+R_{\mu \nu \rho \sigma}R^{\mu \nu \rho \sigma}\,,
\end{equation}
one can avoid the appearance of spurious spin-2 ghosts
\cite{ghost1,ghost2}.  If a scalar field $\phi$ with an exponential
potential $V(\phi)=V_0 e^{-\lambda \phi}$ couples to the GB term
\cite{NO05}, a scaling matter era can be followed by a late-time de
Sitter solution for the exponential GB coupling $F(\phi) \propto
e^{\mu \phi}$ with $\mu>\lambda$ \cite{KMota1,TS07}.  While this GB
coupling is well motivated by low-energy effective string theory
\cite{Gas} the joint likelihood analysis using observational data of
big bang nucleosynthesis, large scale structure, and baryon acoustic
oscillations disfavors such a model, provided that it aims to account
for dark energy \cite{KMota2}. In addition the energy contribution
coming from the GB term needs to be strongly suppressed for
consistency with solar-system experiments \cite{Amen,Thomas}. 
This cannot be compatible with the requirement of cosmic acceleration today (see also
Ref.~\cite{Davis}),  at least in the presence of a kinetic term for the
scalar field and some forms of the potential.  The instability of
tensor perturbations is also present in those models during the epoch
of cosmic acceleration \cite{Kawai,TS07,Guo}.

However, it is possible to explain the late-time cosmic acceleration
for the modified gravity scenario in which the Lagrangian density is
given by ${\cal L}=R+f(\GB)$, where $f(\GB)$ is an arbitrary function
in terms of $\GB$ \cite{fGO}, provided the function $f$ satisfies some
conditions \cite{DT08}. This is equivalent to a theory 
with a scalar field coupled to the GB term in the absence of 
a kinetic term \cite{fGO,Soti07,Uddin}.
A number of $f(\GB)$ models that have a
matter era followed by a de Sitter (dS) attractor have been proposed
in Ref.~\cite{DT08} (see also Refs.~\cite{Cognola,Li,Zhou}).  These
models can be also consistent with local gravity constraints for a
wide range of parameter space \cite{DT09}.

In order to test the cosmological viability of $f(\GB)$ dark energy
models, it is important to study cosmological
perturbations responsible for structure formation.
In Ref.~\cite{Li} the evolution of density perturbations has
been discussed for non-relativistic matter with an equation
of state $w=0$, under the approximation that the background 
cosmological evolution mimics that of the $\Lambda$CDM model.
In this paper we derive the equation for density perturbations 
with a general constant equation of state $w$.
Therefore our analysis includes the perturbations 
in radiation ($w=1/3$) as well as those in non-relativistic matter.
Moreover we use concrete $f(\GB)$ models that satisfy
both local gravity constraints and cosmological constraints at the 
background level.
This is particularly important when we discuss the evolution 
of perturbations at late times, because the deviation from 
the $\Lambda$CDM model can be significant.

We will show that, in the Universe dominated by a single perfect fluid
with an equation of state  parameter $w$,  
the perturbations in the fluid exhibit violent instabilities for
$w>-1/2$ in the small-scale limit.
This is associated with a negative speed squared $c_s^2$ of
one eigenvector-mode.
The perturbations of non-relativistic matter as
well as radiation, at some scale, will start to grow exponentially
during the matter/radiation domination, unless the deviation from GR
is very small.
In GR, this same mode does not exist, 
so that there is no smooth limit from one theory to the other.

This paper is organized as follows.
In Sec.~\ref{sec:tm} we review $f(\GB)$ models that 
satisfy cosmological constraints at the background level
as well as local gravity constraints.
Sec.~\ref{sec:pgwm} is devoted to the analysis of cosmological
perturbations in $f(\GB)$ gravity 
in the presence of a perfect fluid with the
equation of state $w$. In this section we discuss the presence
of an instability at small scales due to a negative speed squared 
for the propagating mode. In Sec.~\ref{sec:mg}
we analyze more in detail for perturbations in non-relativistic matter 
in order to describe the growth of large-scale structure.
We shall numerically integrate the perturbation equations for concrete 
$f(\GB)$ models and estimate how much deviation from GR can be 
allowed by the observations of galaxy clustering in the linear regime.
We conclude in Sec.~\ref{sec:c}.

\section{Dark energy models based on $f(\GB)$ gravity }
\label{sec:tm}

Let us start with the following action
\begin{equation}
\label{eq:ac1}
S=\frac{1}{16\pi G}\int {\rm d}^4x\sqrt{-g_M}\, [R+f(\GB)]+S_{m}\,,
\end{equation}
where $G$ is a bare gravitational constant and $g_M$ 
is the determinant for the space-time metric $g_{\mu \nu}$.
For the matter action $S_m$
we shall consider a perfect fluid whose equation of state
$w=p_m/\rho_m$ is strictly constant, where $p_m$ and $\rho_m$ are the
pressure and the energy density respectively. 
Taking the variation of the action (\ref{eq:ac1}) with respect 
to $g_{\mu \nu}$, we obtain the field equation
\begin{eqnarray}
\label{geeq}
& &G_{\mu \nu}+8[ R_{\mu \rho \nu \sigma} +R_{\rho \nu} g_{\sigma \mu}
-R_{\rho \sigma} g_{\nu \mu} -R_{\mu \nu} g_{\sigma \rho} \nonumber \\
& &+R_{\mu \sigma} g_{\nu \rho}+(R/2)(g_{\mu \nu} g_{\sigma \rho}
-g_{\mu \sigma} g_{\nu \rho})] \nabla^{\rho} \nabla^{\sigma} f_{,\GB} 
\nonumber \\
& &+(\GB f_{,\GB}-f) g_{\mu \nu}=8\pi G T_{\mu \nu}\,,
\end{eqnarray}
where $f_{,\GB} \equiv \partial f/\partial \GB$, 
$G_{\mu \nu} \equiv R_{\mu \nu}-(1/2)R g_{\mu \nu}$ is the Einstein-tensor, 
and $T_{\mu \nu}$ is the energy momentum tensor of matter.

For the flat Friedmann-Lema\^{i}tre-Robertson-Walker (FLRW) background 
with a scale factor $a$, we obtain the following dynamical equation
\begin{equation}
\label{eq:back}
3H^2=\GB f_{,\GB}-f-24H^3 \dot{f_{,\GB}}
+8\pi G \rho_m\,,
\end{equation}
where $H \equiv \dot{a}/a$, a dot represents a time derivative in
terms of cosmic time $t$, 
and the GB term is given by 
\begin{equation}
\GB=24H^2(H^2+\dot{H})=-12H^4 (1+3w_{\rm eff})\,.
\label{GBdef}
\end{equation}
Here $w_{\rm eff}$ is an effective equation of state defined by 
\begin{equation}
w_{\rm eff} \equiv -1-\frac{2\dot{H}}{3H^2}\,.
\end{equation}
The matter energy density $\rho_m$ satisfies the standard continuity
equation, 
\begin{equation}
\dot{\rho}_m+3H(1+w)\rho_m=0\,,
\end{equation}
which has the solution $\rho_m \propto a^{-3(1+w)}$
for constant $w$.

It is possible to realize a late-time cosmic acceleration by the
existence of a de Sitter (dS) point that satisfies the condition
$3H_1^2=\GB_1f_{\GB}(\GB_1)-f(\GB_1)$, where $H_1$ and ${\cal G}_1$
are the Hubble parameter and the GB term at the dS point respectively.
The condition,
\begin{equation}
0<H_1^6 f_{,\GB \GB} (H_1)<1/384\,,
\label{desta}
\end{equation}
is required from the stability of the dS point \cite{DT08}.  We have $\GB<0$ and
$\dot{\GB}>0$ during both radiation and matter domination. However the GB
term changes its sign from negative to positive during the transition
from the matter era ($\GB=-12H^4$) to the dS epoch ($\GB=24H^4$).  For
the existence of standard radiation and matter eras we require that
$f_{,\GB \GB} \equiv \partial^2 f/\partial \GB^2>0$ for $\GB \le
\GB_1$ \cite{DT08}. Since the term $24H^3 \dot{f_{,\GB}}$ in Eq.~(\ref{eq:back})
is of the order of $H^8f_{,\GB \GB}$, this is suppressed relative to 
$3H^2$ for $H^6 f_{,\GB \GB} \ll 1$ during the radiation and matter
domination. In order for this condition to hold, we require
that $f_{,\GB \GB}$ approaches $+0$ in the limit $|\GB| \to \infty$. 
Recall that even around the de Sitter point the condition 
$H^6 f_{,\GB \GB} \ll 1$ is satisfied from Eq.~(\ref{desta}).

A couple of representative models that can satisfy these
conditions are \cite{DT08}
\begin{eqnarray}
{\rm (A)}~f (\GB) &=& \lambda \frac{\GB}{\sqrt{\GB_*}}\,{\rm arctan}\! 
\left( \frac{\GB}{\GB_*} \right)
-\frac{1}{2}\lambda \sqrt{\GB_*}\,{\rm ln} \!
\left(1+\frac{\GB^2}{\GB_*^2} \right) \nonumber \\
& &-\alpha \lambda \sqrt{\GB_*}\,,
\label{model}
\\
{\rm (B)}~f (\GB) &=&
\lambda \frac{\GB}{\sqrt{\GB_*}}\,{\rm arctan}
\left( \frac{\GB}{\GB_*} \right)
-\alpha \lambda \sqrt{\GB_*}\,,
\label{model2}
\end{eqnarray}
where $\alpha$, $\lambda$ and $\GB_*$ are positive constants.  The
second derivatives of $f$ in terms of $\GB$ for the models (A) and (B) 
are $f_{,\GB\GB}=\lambda/[\GB_*^{3/2} (1+\GB^2/\GB_*^2)]$
and $f_{,\GB\GB}=2\lambda/[\GB_*^{3/2} (1+\GB^2/\GB_*^2)^2]$, 
respectively (both of which are positive for all $\GB$).

The quantity defined by 
\begin{equation}
\xi \equiv f_{,\GB}\,,
\end{equation}
is constant for the $\Lambda$CDM model, $f(\GB)=-2\Lambda+c\,\GB$
(here we have included the linear term $c\,\GB$ because this also 
gives rise to the equations of motion same as those in 
the $\Lambda$CDM model).
In order to discuss cosmological perturbations in the next section, 
it is convenient to introduce the following quantity
\begin{eqnarray}
\label{mudef}
\mu &\equiv& H \dot{\xi}=H\dot{\GB}f_{,\GB\GB} \nonumber \\
&=& 72H^6f_{,\GB \GB} \left[ (1+w_{\rm eff})(1+3w_{\rm eff})-w_{\rm eff}'/2
\right],
\end{eqnarray}
where a prime represents a derivative with respect to $N=\ln a$.
This quantity characterizes the deviation from the $\Lambda$CDM model.
During the radiation and matter domination one has 
$\mu=192 H^6 f_{,\GB \GB}$ and 
$\mu=72 H^6 f_{,\GB \GB}$, respectively, whereas
at the de Sitter attractor $\mu=0$.

\begin{figure}[ht]
\includegraphics[height=3.2in,width=3.4in]{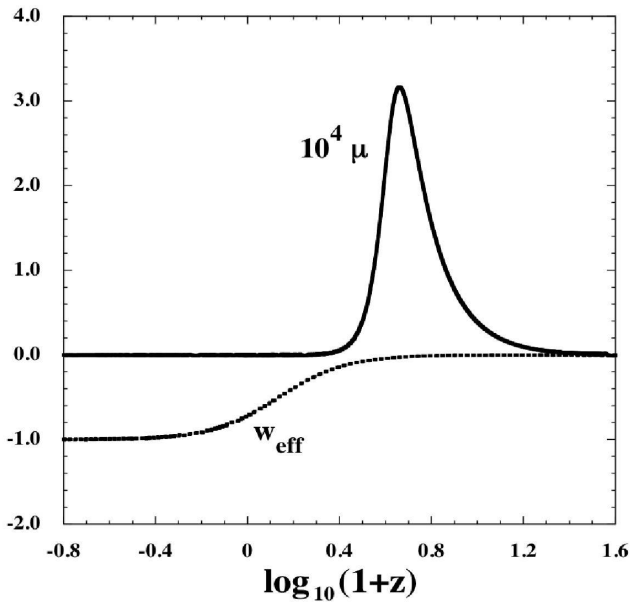}
\caption{
Evolution of $\mu$ (multiplied by $10^4$) and $w_{\rm eff}$
versus the redshift $z=a_0/a-1$ for the model (\ref{model})
with parameters $\alpha=100$ and $\lambda=3 \times 10^{-4}$.
The initial conditions are chosen to be $x=-1.499985$, 
$y=20$, and $\Omega_m=0.99999$ (see Appendix B
for the definition of $x$ and $y$).}
\label{muplot}
\end{figure}

In Fig.~\ref{muplot} we plot the evolution of $\mu$ and $w_{\rm eff}$
in the model (A) for $\alpha=100$ and $\lambda=3 \times 10^{-4}$.  In
this case the quantity $\mu$ is much smaller than unity in the deep
matter era ($w_{\rm eff} \simeq 0$) and it grows to the order of
$10^{-4}$ prior to the accelerated epoch. This is followed by the
decrease of $\mu$ toward 0 with small oscillations, as the solution
approaches the de Sitter attractor with $w_{\rm eff}=-1$.  For smaller
$\alpha$ and larger $\lambda$, it is also possible to realize larger
maximum values of $\mu$ such as $\mu_{\rm max} \gtrsim 0.1$.  The
qualitative behavior shown in Fig.~\ref{muplot} is generic for viable
$f(\GB)$ models at the background level.

\section{Cosmological perturbations}\label{sec:pgwm}

In order to study cosmological perturbations in $f(\GB)$ gravity we
introduce a perturbed metric with 4 scalar perturbations $\alpha$,
$\beta$, $\phi$ and $\gamma$ about a spatially flat FLRW cosmological
background, \cite{cosmoreview}
\begin{eqnarray}
\de s^2 &=& -(1+2\alpha)\,\de t^2-2a\beta_i\,\de t\,\de x^i 
\nonumber \\
& &+a(t)^2 \left[ (1+2\phi)\delta_{ij}
+2\partial_i\partial_j\gamma \right]
\de x^i\de x^j\,.
\label{eq:met1}
\end{eqnarray}
Let us decompose the energy-momentum tensor $T^{\mu}_{\nu}$ into
background and perturbed parts, i.e.  $T^0_0=-(\rho_m+\delta \rho_m)$
and $T^0_{\alpha}=-\rho_m v_{,\alpha}$, where $v$ is 
a velocity potential.
We define the gauge-invariant matter density perturbation
$\delta_m$, as
\begin{equation}
\label{eq:del1}
\delta_m \equiv \frac{\delta\rho_m}{\rho_m}
+\frac{\dot\rho_m}{\rho_m}v\,.
\end{equation}

We also introduce two gauge-invariant combinations
\begin{eqnarray}
\Phi_1 &\equiv& \phi+Hv\,, \\
\Phi_2 &\equiv& \delta \xi
+\dot{\xi}v\,,
\end{eqnarray}
where $\delta \xi$ is the perturbation of the quantity $\xi=f_{,\GB}$.

\subsection{Perturbation equations}

Following  a similar procedure to the one developed recently in
Ref.~\cite{DeSuyama}, one can show that the dynamics of cosmological
perturbations in $f(\GB)$ gravity in the presence of a perfect fluid
with an equation of state $w$ reduces to that of two propagating
fields $\Phi_1$ and $\Phi_2$ defined by the following perturbed action
\begin{eqnarray}
\label{eq:actP}
\delta S &=& \int \de^4x[A_1\dot\Phi_1^2
+2A_2\dot\Phi_1\dot\Phi_2+A_3\dot\Phi_2^2
-g_1(\vec\nabla\Phi_1)^2 \nonumber \\
& &-2g_2\vec\nabla\Phi_1\cdot\vec\nabla\Phi_2-g_3(\vec\nabla\Phi_2)^2
+B(\dot\Phi_2\Phi_1-\dot\Phi_1\Phi_2) \nonumber \\
& &-m_3\Phi_2^2-2m_2\Phi_1\Phi_2]\,,
\end{eqnarray}
where $A_i, g_i, B, m_i$ are time-dependent coefficients whose
explicit forms are given in Appendix A.  We also have the following
relation
\begin{eqnarray}
\alpha+\dot v &=& 
\frac{1+4\mu}{H(1+6\mu)}\,\dot\Phi_1
+\frac{2H}{1+6\mu}\,\dot\Phi_2
-\frac{2H^2}{1+6\mu}\,\Phi_2 \nonumber \\
&=& -\frac{w}{1+w}\delta_m\,,
\label{alphadot}
\end{eqnarray}
where $\mu$ is defined in Eq.~(\ref{mudef}).

{}From the action (\ref{eq:actP}) we obtain the perturbation equations
in Fourier space
\begin{eqnarray}
\label{eq:EQp1}
& &\frac{\de}{\de t}(A_1\dot\Phi_1+A_2\dot\Phi_2)
+g_1k^2\Phi_1+g_2k^2 \Phi_2 \nonumber \\
& &+2m_2\Phi_2-B\dot\Phi_2=0\,,\\
& &\frac{\de}{\de t}(A_2\dot\Phi_1+A_3\dot\Phi_2)
+g_2k^2 \Phi_1+g_3 k^2\Phi_2 \nonumber \\
& & +m_3\Phi_2+B\dot\Phi_1=0\,,
\label{eq:EQp2}
\end{eqnarray}
where $k$ is a comoving wavenumber.

\subsection{Instability at small scales}

Let us show the presence of a small-scale instability in $f(\GB)$
gravity associated with a negative speed squared of one propagating
mode.  This instability appears at large redshifts for the models that
look like GR ($\mu\ll1$) at early times.  For sufficiently small
scales it is easy to show that the highest derivative terms prevail
over any other terms in the differential equations. Then one can
approximately find a harmonic oscillator-like dispersion relation in
the follow way.

For large $k$, one can look for an approximate solution in 
the following way. The dominant
contribution to the equation of motion will be in the form
\begin{equation}
\label{eq:pip}
A\ddot{\vec\Phi}-g\nabla^2\vec\Phi
\approx 0\,,
\end{equation}
where $A$ is the $2 \times 2$ symmetric matrix whose diagonal elements
are $A_1$ and $A_3$ and non-diagonal element is $A_2$. Along the same
lines one defines the matrix $g$ with the elements $g_1$, $g_2$, and
$g_3$. Introducing the time-dependence $\Phi_j \propto\exp(i \omega t)$
with $j=1, 2$
in Fourier space, it follows that
\begin{equation}
\label{eq:pip2}
(-\omega^2 A+k^2g)\,\vec{\Phi}
\approx 0\,,
\end{equation}
which has solutions for some values of $\omega^2$. 
This approximation tends to be more accurate for 
larger $k$. These expressions give the dispersion relation 
for the propagating modes under consideration.

Non-zero solutions for $\vec{\Phi}$ exist provided 
that the following relation holds
\begin{equation}
\label{eq:dispR}
\det(\omega^2A-k^2g)=0\,.
\end{equation}
After finding the eigenvalues $\omega^2$, one can proceed to 
look for the eigenvectors which diagonalize the kinetic operator. 
We find that one eigenvector-mode propagates with a speed squared
$c_1^2=w$, as expected, and the other one with a speed squared
\begin{equation}
\label{eq:eqcs2w}
c_2^2=1+\frac{2\dot H}{H^2}+\frac{1+w}{1+4\mu}
\frac{8\pi G\rho_m}{3H^2}\,.
\end{equation}
This coincides the result of the vacuum case found in
Ref.~\cite{DeFelice09} by taking the limit $\rho_m \to 0$.

In the Universe dominated by a single fluid one has $3H^2 \simeq 8\pi
G\rho_m$ and $\dot{H}/H^2 \simeq -3(1+w)/2$.  Under the condition that
$\mu \ll 1$, the speed squared (\ref{eq:eqcs2w}) reduces to
\begin{equation}
c_2^2 \simeq -1-2w\,.
\label{csap}
\end{equation}
This shows the existence of a negative instability for $w>-1/2$.
Hence the perturbations in radiation and non-relativistic matter are
subject to this instability during the radiation and matter
domination, respectively.  In the matter-dominated epoch ($\Omega_m
\simeq 1$) the result (\ref{csap}) agrees with the value $\tilde
c_2^2=1+4\dot{H}/(3H^2)$ that appears as a coefficient of the term
$k^2/a^2$ in Eq.~(47) of Ref.~\cite{Li}.  During the transition from
the matter era to the accelerated epoch $c_2^2$ can be quite different
from $\tilde c_2^2$, because $\Omega_m$ is smaller than 1 and the
quantity $\mu$ is not necessarily negligible relative to 1. 
Therefore the stability at late times must be checked against the
quantity $c_2^2$.

The reason why the values $c_2^2$ and $\tilde c_2^2$ are different can
be understood as follows. Looking at
Eqs.~(47) and~(48) in Ref.~\cite{Li}, 
the modes are not diagonalized, as the
$k^2/a^2$-term still appears for two different fields.  Even if the
coefficient of the $k^2/a^2$ term for one of two equations is
negative, this is not enough to state that there is an instability in
the system.  In other words, even if $g_1$ is negative, the
eigenvalues of $A^{-1}g$ can still be both positive. Another
difference among these two studies is that for the evolution of
perturbations the authors in Ref.~\cite{Li} chose a background
solution that mimics the evolution in the $\Lambda$CDM 
model\footnote{Observations show that the current equation of state 
of dark energy is close to $-1$ to a pretty high redshift,
so that the deviation from the GR background is not large.
Ref.~\cite{Li} used this fact to neglect the $k^2 \eta$ 
contribution relative to the term already in the LHS of Eqs.~(47) for 
the realistic cosmological models one considers, 
for which $\mu \ll1$ or equivalently $|H^6f_{,\GB \GB}|\ll1$.
In this work we make a more general approach 
of computing the diagonalized modes accurately 
by including the contribution coming from the $k^2 \eta$ term.}. 
Strictly speaking, this is not a solution of the Einstein equations.  This can
affect the evolution of the perturbations especially at late times.
In our work we use concrete $f(\GB)$ models to find cosmological
evolution of both the background and the perturbations. 

The Laplacian instability mentioned above appears at any time in the
past at some sufficiently small scales (apart from the epoch of
inflation during which $c_2^2 \approx 1$).  This can place tight
constraints on $f(\GB)$ models.  Let us discuss this more precisely,
without any approximation, for the growth of large-scale structure
during the matter domination.

\section{Growth of matter perturbations}\label{sec:mg}

Let us focus on non-relativistic matter with the equation of state
$w=0$. In order to derive the equation for matter perturbations, we
first combine Eqs.~(\ref{alphadot}), (\ref{eq:EQp1}), (\ref{eq:EQp2})
and finally take the limit $w \to 0$.  {}From Eq.~(\ref{alphadot}) one
can express $\dot{\Phi}_1$ in terms of $\dot{\Phi}_2$, $\Phi_2$ and
$\delta_m$.  Multiplying Eq.~(\ref{eq:EQp1}) by $g_2$ and
Eq.~(\ref{eq:EQp2}) by $g_1$ and subtracting one to the other, we find
an equation which depends on $\Phi_1$ only through its first time
derivative. Combining these two equations and taking the limit $w \to
0$, one reaches the following equation of motion
\begin{equation}
\label{eq:EQX}
\ddot\Phi_2-d_4 \dot{\Phi}_2+\left( d_3+
c_2^2 \frac{k^2}{a^2} \right)\Phi_2
-d_1\dot\delta_m-d_2\delta_m=0\,,
\end{equation}
where 
\begin{eqnarray} 
\hspace*{-2.7em}& &c_2^2=\frac{1+\Omega_m+2x+4\mu (1+2x)}{1+4\mu}, \\
\hspace*{-2.7em}& & d_1=\frac{\mu [1+\Omega_m+2x+4\mu (1+2x)]}{H(1+4\mu)}, \\
\hspace*{-2.7em}& & d_2=\frac{\Omega_m [1+3\Omega_m+4x+4\mu (1+4x)]}{4(1+4\mu)}, \\
\hspace*{-2.7em}& & d_3=H^2 \{ 4x^2+4x+x'-2\mu [4(1-3x^2-2x-x') \nonumber \\
\hspace*{-2.7em}& &~~~+3\Omega_m(1+x)+8\mu (2-2x^2-x')] \}/
[2\mu (1+4\mu)],
 \label{d3} \\
\hspace*{-2.7em}& & d_4=-\frac{3H[1+\Omega_m+2x+4\mu (1+2x)]}{1+4\mu}\,,
\end{eqnarray} 
and $x \equiv \dot{H}/H^2$, $x' \equiv \dot{x}/H$ and $\Omega_m \equiv
8\pi G\rho_m/(3H^2)$.

To find the second dynamical equation we multiply Eq.~(\ref{eq:EQp1})
by $g_3$ and Eq.~(\ref{eq:EQp2}) by $g_2$ and then subtract the two
equations. We then divide it by $g_1g_3-g_2^2$ and differentiate it
with respect to time.  This gives rise to the equation which involves
$\dot\Phi_1$, so that we can replace it with $\delta_m$.  Furthermore
the same equation will contain second and third time-derivatives of
$\Phi_2$, which can be substituted by using Eq.~(\ref{eq:EQX}). By
doing so and taking the limit $w \to 0$, one finds the following
dynamical equation
\begin{equation}
\label{eq:EQD}
\ddot\delta_m-d_5\dot\delta_m-d_6\delta_m
-d_8\dot\Phi_2+\left(d_9+d_7 \frac{k^2}{a^2} \right) \Phi_2=0\,,
\end{equation}
where
\begin{eqnarray} 
\hspace*{-2.7em}& & d_5= -\frac{2H (1-2x \mu+2\mu)}{1+4\mu},\quad
d_6=\frac{3H^2 \Omega_m (1+x)}{1+4\mu},\\
\hspace*{-2.7em}& & d_7=\frac{4H^2 (1+x)}{1+4\mu},\quad
d_8=-\frac{12 H^3 (1+x)}{1+4\mu},\\
\hspace*{-2.7em} & & d_9=\frac{3H^4[4x^2+4x+x'-4\mu (1-3x^2-2x-x')]}
{\mu (1+4\mu)}\,.
\label{d9}
\end{eqnarray}

Note that in GR $\mu=0$ and $\Phi_2=0$.  {}From Eqs.~(\ref{d3}) and
(\ref{d9}) both $d_3$ and $d_9$ diverge in the limit $\mu \to 0$.
Therefore we can solve Eq.~(\ref{eq:EQX}) for $\Phi_2$ and substitute
it into Eq.~(\ref{eq:EQD}).  This results in the following equation
\begin{equation}
\label{delmeq}
\ddot{\delta}_m+C_1 \dot{\delta}_m+C_2 \delta_m=
r \ddot{\Phi}_2+(d_8-r d_4)\dot{\Phi}_2\,,
\end{equation}
where $C_1 \equiv r d_1-d_5$, 
$C_2 \equiv r d_2-d_6$, and
\begin{equation}
r \equiv \frac{M_B^2}{M_A^2}\,, \quad
M_A^2 \equiv d_3+c_2^2\frac{k^2}{a^2}\,,\quad
M_B^2 \equiv d_9+d_7\frac{k^2}{a^2}\,.
\end{equation}
In order to derive analytic solutions let us consider the case $\mu
(k/aH)^2 \ll 1$.  Since we are interested in sub-horizon modes ($k \gg
aH$), the condition $\mu \ll 1$ also follows.  During the matter
domination characterized by $H \simeq 2/(3t)$ and $\Omega_m \simeq 1$
we have
\begin{equation}
C_1 \simeq 2H (1-2\mu)\,,\quad
C_2 \simeq -\frac32 H^2 \left[1+ \frac89
\mu \left( \frac{k}{aH} \right)^2 \right]\,.
\label{C1C2}
\end{equation}
In the GR limit $\mu \to 0$ and $\Phi_2 \to 0$, Eq.~(\ref{delmeq})
reduces to $\ddot\delta_m+2H\dot\delta_m-(3/2)H^2 \delta_m=0$, which
has the growing mode solution $\delta_m \propto t^{2/3}$ in the matter
era.

In the regime $\mu(k/aH)^2 \ll 1$ the growth rate of $\delta_m$ gets
larger than that in GR because of the presence of the $(8/9)\mu
(k/aH)^2$ term in Eq.~(\ref{C1C2}).  For sub-horizon modes ($k \gg
aH$) this effect is more important than the reduction of the friction
term $C_1$ induced by $2\mu$.

The quantity $\mu(k/aH)^2$ can grow to the order of 1 by the present
epoch, depending on the wavenumber $k$.  In this case the growth of
matter perturbations is significantly different from that in
GR. During the matter-dominated epoch one has $d_3 \simeq
3H^2/(2\mu)>0$ and $c_2^2 \simeq -1$, so that the mass term $M_A^2
\simeq 3H^2/(2\mu)-k^2/a^2$ changes its sign from positive to negative
at $\mu (k/aH)^2=3/2$.  This leads to a negative instability for the
perturbation $\Phi_2$ through Eq.~(\ref{eq:EQX}).  The evolution of
the mass term $M_B^2$ during the matter era is given by $M_B^2 \simeq
H^2(9H^2/\mu-2k^2/a^2)$, which changes from positive to negative at
$\mu (k/aH)^2=9/2$.  Thus the onset of the negative instability can be
characterized by the condition
\begin{equation}
\mu \approx \left( aH/k \right)^2\,.
\label{mucri}
\end{equation}

In the regime $\mu (k/aH)^2 \gg 1$ one can approximate $M_A^2 \simeq
c_2^2k^2/a^2$ and $M_B^2 \simeq d_7k^2/a^2$, which results in the
positive mass ratio $r \simeq 2H^2/(1+8\mu)$.  Then
Eq.~(\ref{delmeq}) reduces to
\begin{equation}
\label{delmda}
\ddot\delta_m+2H\dot\delta_m+\frac{H^2}{2(1+8\mu)}
\delta_m=\frac{2H^2}{1+8\mu}\ddot{\Phi}_2\,.
\end{equation}
Here we have not used the approximation $\mu \ll 1$.  Notice that the
coefficient in front of $\delta_m$ is positive and hence this term
does not lead to the growth of $\delta_m$.  However, the rapid growth
of $\Phi_2$ induced by the negative $c_2^2$ works as a source term for
the amplification of $\delta_m$ in Eq.~(\ref{delmda}).  We have
$c_2^2=1$ and $d_7=4H^2$ at the late-time dS point, which means that
both $c_2^2$ and $d_7$ change signs from negative to positive during
the transition from the matter era to the accelerated epoch. Hence we
can expect that the growth of matter perturbations ends before
reaching the dS attractor.
 
\begin{figure}[ht]
\includegraphics[height=3.2in,width=3.4in]{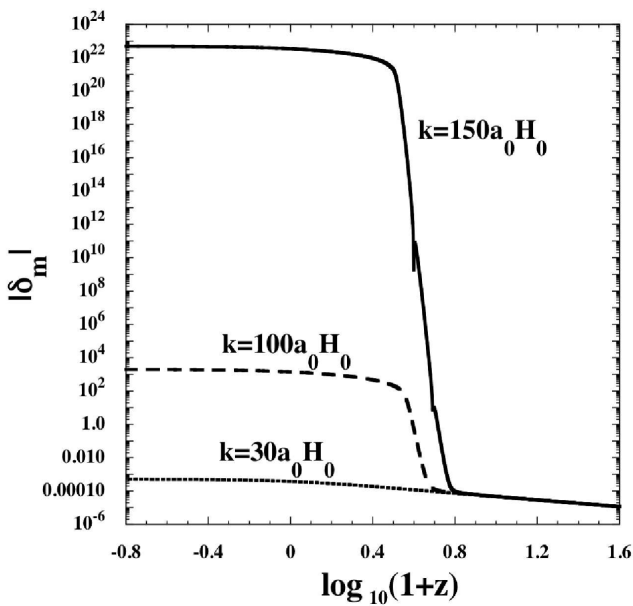}
\caption{
Evolution of $\delta_m$ versus the redshift $z=a_0/a-1$
for the model (\ref{model})
with the same model parameters as given 
in Fig.~\ref{muplot}.
We choose three different wave-numbers: 
(i) $k=150a_0H_0$, (ii) $k=100a_0H_0$, and 
(iii) $k=30a_0H_0$.
The initial conditions are $x=-1.499985$, 
$y=20$, $\Omega_m=0.99999$, and 
$\delta_m=\dot{\delta}_m/H=10^{-5}$, 
$\Phi=\dot{\Phi}=0$.}
\label{delta}
\end{figure}

In Fig.~\ref{delta} the evolution of matter perturbations is plotted
for the model (\ref{model}) with parameters $\alpha=100$ and
$\lambda=3 \times 10^{-4}$ (see Appendix B for the detail of 
numerical integration).
In this case the quantity $\mu$ reaches the maximum value 
$\mu_{\rm max}=3 \times 10^{-4}$ around the redshift $z=3.6$ 
(see Fig.~\ref{muplot}). Using the criterion (\ref{mucri}), the 
perturbation with $k \approx 60 aH$ is about to enter the negative 
instability region. The quantity $\mu$ decreases rapidly after
it reaches the maximum, whereas $aH$ at $z=3.6$ is not 
much different from $a_0H_0$ today ($z=0$).
Hence one can estimate that the modes with 
$k \lesssim 60 a_0H_0$ are hardly affected by the negative instability.
In the numerical simulation of Fig.~\ref{delta} this can be confirmed 
for the mode $k=30a_0H_0$.
Meanwhile Fig.~\ref{delta} shows that the modes with 
$k\gtrsim 100a_0H_0$ exhibit violent negative instabilities.
Note that the apparent discontinuous behavior seen in Fig.~\ref{delta} 
for the mode $k=150a_0 H_0$ comes from the fact that 
$\delta_m$ temporally becomes negative.

The wave-numbers relevant to the observed galaxy power spectrum in the
linear regime corresponds to $30a_0H_0 \lesssim k \lesssim 600a_0H_0$
(i.e. $0.01\,h$\,Mpc$^{-1} \lesssim k \lesssim 0.2$\,h\,Mpc$^{-1}$).
For the model parameters used in Fig.~\ref{delta}, the resulting
matter power spectrum is certainly ruled out from the observations of
large scale structure. In Fig.~\ref{delta} we have chosen the initial
conditions $\Phi=\dot{\Phi}=0$ as a minimal case, but non-zero initial
values of $\Phi$ and $\dot{\Phi}$ lead to even larger amplitude of
$\delta_m$.  We also note that, irrespective of the forms of $f(\GB)$
models discussed in Sec.~\ref{sec:tm}, the behavior of perturbations
is similar to that discussed above.

The only way to avoid this negative instability is to make the
parameter $\mu$ as small as possible by changing model parameters, so
that the modes relevant to the matter power spectrum never reach the
regime $\mu (k/aH)^2={\cal O}(1)$. If we take the smallest scale $k
\approx 600a_0H_0$ of the linear matter power spectrum, the condition
under which the negative instability can be avoided translates into
\begin{equation}
\mu_{\rm max} \lesssim 10^{-6}\,,
\label{mumax}
\end{equation}
where we have used the approximation $aH \approx a_0H_0$ at
$\mu=\mu_{\rm max}$.  Hence the deviation from the $\Lambda$CDM model
is constrained to be very small. Nonetheless, even by introducing
by hand this effective cutoff for the wavelength due to the experimental
apparatus we use to observe data, the theory does possess an
ultra-violet (UV) instability, no matter how small but non-zero $\mu$
is. In this case perturbation theory at some small scale will break
down, during anytime in the past up to the dark energy domination.
Therefore, these theories cannot be studied by using
perturbation theory, and in general, one should expect strong
dynamical deviations from GR, as the background is not 
trustable any longer.

Furthermore we wish to stress that the negative instability
cannot be avoided as we go to smaller scales. In order to avoid
violent growth in the non-linear regime of the matter power spectrum
($k \gtrsim 600a_0H_0$), the constraint on $\mu_{\rm max}$ becomes
even severer than the one given in Eq.~(\ref{mumax}).  Moreover the
growth rate of matter perturbations gets enormously large for
increasing $k$.  The point is that we can always find the wavenumber
$k$ satisfying $\mu(k/aH)^2 \approx 1$ even for very small values of
$\mu$.  This property also persists for the perturbations in
radiation.  Since the quantity $\mu$ during the radiation era is
suppressed relative to that during the matter era, the scales of
instabilities of radiation perturbations are much smaller than those
of matter perturbations. The only way to consistently remove this
UV instability is to set $\mu$ identically equal to zero, that is, 
the gravitational theory exactly reduces to GR.

\section{Conclusions}\label{sec:c}

In this paper we have studied cosmological perturbations in $f(\GB)$
gravity, in the presence of a perfect fluid with a constant equation
of state $w$.  In the Universe dominated by a single fluid with $w >
-1/2$, we have shown the presence of an instability associated with a
negative speed squared of one eigenvector-mode.  Hence the
perturbations in radiation and non-relativistic matter are affected by
this instability during the radiation and matter domination,
respectively.  Our results are more general than those given in
Ref.~\cite{Li} in the sense that we have considered a general equation
of state $w$ and that we have not assumed the $\Lambda$CDM-like
background evolution.

A useful quantity that characterizes the deviation from the
$\Lambda$CDM model is $\mu=H \dot{f}_{,\GB}$.  In the limit that $\mu
\to 0$ (i.e. the $\Lambda$CDM model) one can avoid the appearance of
the negative instability.  If $\mu \ne 0$, the instability of
perturbations appears for $\mu \gtrsim (aH/k)^2$.  Even for tiny
values of $\mu$ much smaller than 1, there are small scale modes that
satisfy this condition.  We have studied the evolution of
non-relativistic matter perturbations numerically and confirmed that
the perturbations are strongly amplified once they enter the regime
$\mu \gtrsim (aH/k)^2$. {}From the requirement that the matter power
spectrum in the linear regime is not affected by this violent
instability, we have found that the maximum value of the deviation
parameter is constrained to be $\mu_{\rm max} \lesssim 10^{-6}$.

Nonetheless the UV limit of this theory remains unsatisfactory,
as perturbation theory would break down eventually at some scale,
and the background is not under control any longer at these
scales. When this happens, there is not even an easy way, for the
cosmological background of this theory to be checked against
observations. It is mostly this UV unpredictable behavior 
which sets the strongest bound. This feature will always
remain unless one sets $\mu$ identically to 0 at any time, as in
this case the theory reduces to GR and the unstable eigenvector-mode
automatically disappears. In this sense we believe that this theory
is ruled out by our analysis, which explicitly shows 
the existence of eigenmodes with negative squared
speed in the past.

While we have focused on linear perturbations, non-linear effects
become important once $\delta_m$ grows to the order of 1. 
It may be of interest to see whether such non-linear effects strengthen or
weaken the growth of perturbations.

\section*{ACKNOWLEDGEMENTS}
We thank M.~Sandstad for discussions and comments. 
The	work of A.\,D and S.\,T. was supported by the Grant-in-Aid 
for Scientific Research Fund 
of the JSPS Nos.~09314 and 30318802.
S.\,T. also thanks financial support for the
Grant-in-Aid for Scientific Research on Innovative 
Areas (No.~21111006).

\begin{widetext} 
\section*{Appendix A: Coefficients of the action (\ref{eq:actP})}
Here we present the coefficients appearing in the action (\ref{eq:actP}): 
\begin{eqnarray} 
& & A_1={\frac { ( 12\, H^{5} \dot\xi^{3}w+3\, H ^{4}w \dot\xi^{2}+16\, H^{2} \dot\xi ^{2}\pi \,G \rho_m w+16\, 
H ^{2} \dot\xi ^{2}\pi \,G \rho_m   +8\,\pi \,G\rho_m   H  \dot\xi  +8\,\pi \,G \rho_m   wH  \dot\xi  +w\rho_m   \pi \,G+\pi \,G\rho_m  )  
a ^{3}}{2\pi Gw ( 1+6\, \dot\xi H ) ^{2} H ^{2}}},\nonumber \\
\\
& & A_2={\frac { ( -12\, H ^{4}w \dot\xi ^{2}-3\,w \dot\xi  H ^{3}+8\,\pi \,G\rho_m H \dot\xi +8\,\pi \,G\rho_m wH \dot\xi +2\,\pi \,
G\rho_m +2\,w\rho_m \pi \,G )  a ^{3}}{2wG\pi( 1+6\, \dot\xi H) ^{2} }},\\
& & A_3={\frac { (12\,w \dot\xi  H ^{3}+3\,w H ^{2}+4\,w\rho_m \pi \,G+ 4\,\pi \,G\rho_m )  a ^{3} H ^{2}}{2wG\pi( 1+6\, \dot\xi H ) ^{2} }},\\
& & g_1=\frac1{2\pi G ( 1+6\, \dot\xi H  ) ^{2} H ^{2}}( 12\, H ^{5} \dot\xi ^{3}+3\, \dot\xi ^{2} H ^{4}+24\, H ^{3} \dot\xi ^{3}\dot H +24\, 
H ^{2} \dot\xi ^{2}\pi \,G \rho_m +6\, H ^{2} \dot H  \dot\xi ^{2}+24\, H ^{2} \dot\xi ^{2} \pi \,G\rho_m w \nonumber \\
& & \qquad+8\,\pi \,G\rho_m H \dot\xi +8\, \pi \,G\rho_m wH \dot\xi +w\rho_m \pi \,G+\pi \,G \rho_m ) a,\\
& & g_2={\frac { ( -12\, \dot\xi ^{2} H ^{4}-3\, \dot\xi  H ^{3}-24\, H ^{2} \dot H  \dot\xi ^{2}-6\, \dot\xi H \dot H +2\,\pi \,G\rho_m +2\,
w\rho_m \pi \,G ) a }{2\pi G ( 1+6\, \dot\xi H  ) ^{2}}}, \\
& & g_3={\frac {3a H ^{2} ( 4\, \dot\xi  H ^{3}+ H ^{2}+8\, \dot\xi H \dot H +2\,\dot H +4\,w\rho_m \pi \, G+4\,\pi \,G\rho_m ) }{2\pi \,
G ( 1+6\, \dot\xi H ) ^{2}}}, \\
& & B={\frac { (1+ 4\, \dot\xi H ) ( -6\,w \dot\xi  H ^{3}+18\,H  \dot H w\dot\xi +3\, \dot H w+4\,w\rho_m \pi \,G+4\,\pi \,G\rho_m ) 
H  a ^{3}}{4wG\pi( 1+ 6\, \dot\xi H) ^{2} }}, \\
& & m_2=-\dot{B}/2\,,\\
& & m_3=-\frac1{4( 1+6\, \dot\xi H ) ^{3} \dot\xi wG \pi}\left[a ^{3}H  \left( -12\,w \dot H  H ^{2}-3\,wH \ddot H +24\, H ^{5}w\dot\xi 
+ 432\, H ^{7}w \dot\xi ^{3}+192\, H ^{6}w \dot\xi ^{2}\right.\right. \nonumber \\
& &\quad+\left.192\,H  \dot\xi {w}^{2} \rho_m ^{2}{\pi }^ {2}{G}^{2}+72\,\pi \,G\rho_m {w}^{2} \dot\xi  H ^{3}+240\, H ^{2} \dot\xi ^{2} \dot 
H \pi \,G\rho_m +72\,H  \dot\xi  \dot H \pi \,G \rho_m +384\,H  \dot\xi w \rho_m ^{2}{\pi }^{2}{G}^{2}+144\,\pi \,G\rho_m {w}^{2} \dot\xi ^{2} H ^{4}\right. \nonumber \\
& & \quad+\left.96\,\pi \,G\rho_m w \dot\xi ^{2} H ^{4}+80\,\pi \,G\rho_m w \dot\xi H ^{3}-432\, H ^{3} \dot\xi ^{3} \dot H ^{2}w- 252\, 
H ^{3} \dot\xi ^{2} \ddot H w-288\, H ^{2} \dot\xi ^{2} \dot H ^{2}w-48\, H ^{2} \dot\xi  \ddot H w-72\,H  \dot\xi  \dot H ^{2}w\right. \nonumber \\
& & \quad-\left.864 \, H ^{5} \dot H w \dot\xi ^{3}-132\, H ^{3} \dot H w\dot\xi -552\, H ^{4} \dot H w \dot\xi ^{2}-432\, H ^{4} 
\dot\xi ^{3} \ddot H w+192\,H  \dot\xi {\pi }^{2}{G}^{2} \rho_m ^{2}-48\,\pi \,G\rho_m  \dot\xi ^{2} H ^{4}+8\,\pi \,G\rho_m  \dot\xi  H ^ {3}\right. \nonumber \\
& &\quad+\left.\left.672\, H ^{2} \dot\xi ^{2} \dot H w\rho_m \pi \,G+432\, H ^{2} \dot\xi ^{2} \dot H {w}^{2}\rho_m \pi \,G+72\,H  
\dot\xi  \dot H {w}^{2}\rho_m \pi \,G+144\,H  \dot\xi  \dot H w\rho_m \pi \,G-6\,w \dot H ^{2} \right)\right]\,.
\end{eqnarray} 
%

\section*{Appendix B: Numerical integration of dynamical equations}

In addition to the dimensionless variables $x=\dot{H}/H^2$ and 
$\Omega_m=8\pi G \rho_m/(3H^2)$ we introduce another variable 
$y \equiv H/H_*$, where $H_*$ is a constant related to 
$\GB_*$ (a typical scale of the GB term for dark energy) via
$H_*=\GB_*^{1/4}$. 
Then the background equations can be expressed as
\begin{eqnarray} 
& &x'=-4x^2-4x+\frac{1}{24^2H^6 f_{,\GB \GB}}
\left[ \frac{\GB f_{,\GB}-f}{H^2}
-3(1-\Omega_m) \right]\,, 
\label{beeq1} \\
& &y'=xy\,,
\label{beeq2} \\
& &\Omega_m'=-(3+2x)\Omega_m\,.
\label{beeq3} 
\end{eqnarray} 
The quantities $H^6 f_{,\GB \GB}$ and $(\GB f_{,\GB}-f)/H^2$ can be 
expressed by $x$ and $y$ once the model is specified.
Introducing the following quantity
\begin{equation}
\Phi \equiv H^2\Phi_2\,,
\end{equation}
the perturbation equations (\ref{eq:EQX}) and (\ref{eq:EQD}) 
can be written as 
\begin{eqnarray} 
& &\Phi''=\left( \frac{d_4}{H}+3x \right)\Phi'
-\left[ \frac{d_3}{H^2}+2\frac{d_4}{H}x
+2x^2-2x'+c_2^2 \left( \frac{k}{aH} \right)^2
\right]\Phi+d_1 H \delta_m'+d_2 \delta_m \,,
\label{pereq1}\\
& &\delta_m''=\left( \frac{d_5}{H}-x \right)\delta_m'
+\frac{d_6}{H^2}\delta_m+\frac{d_8}{H^3}\Phi'-
\left[ \frac{d_9}{H^4}+\frac{d_7}{H^2}
\left( \frac{k}{aH} \right)^2
+2\frac{d_8}{H^3}x \right] \Phi\,.
\label{pereq2}
\end{eqnarray}
Numerically we solve these equations together with 
the background equations (\ref{beeq1})-(\ref{beeq3}).
 
\end{widetext}


\end{document}